\documentclass[onecolumn]{article}
\usepackage{xcolor}
\usepackage[numbers]{natbib}
\usepackage{hyperref}
\usepackage{cleveref}
\usepackage{graphicx,subfigure}
\RequirePackage{amssymb}

\begin{document}

\title{Tracking the time course of reproduction number and lockdown's effect during SARS-CoV-2 epidemic: nonparametric estimation}

\author{G. Pillonetto$^\ast$, M. Bisiacco$^\ast$, G. Pal\`u$^{\dag}$ and C. Cobelli$^{\ast \circ}$ \vspace{0.2cm} \\ 
$^\ast$ Department of Information Engineering, University of Padova, Italy\\
$^\dag$ Department of Molecular Medicine, University of Padova, Italy\\
and Azienda Zero, Regione Veneto, Italy\\
$^\circ$ Member of Consiglio Superiore di Sanit\`a, Italian Ministry of Health}

\maketitle
\vspace{-0.5cm}
\begin{abstract}
Understanding the SARS-CoV-2 dynamics has been subject of intense research in the last months. In particular, accurate modeling of 
lockdown effects on epidemic evolution is a key issue in order e.g. to inform health-care decisions on emergency management. In this regard, the compartmental and spatial models so far proposed use parametric descriptions of the contact rate, often assuming a time-invariant effect of the lockdown. In this paper we show that these assumptions may lead to erroneous evaluations on the ongoing pandemic. Thus, we develop a new class of nonparametric compartmental models able to describe how the impact of the lockdown varies in time. Our estimation strategy does not require significant Bayes prior information and exploits regularization theory. Hospitalized data are mapped into an infinite-dimensional space, hence obtaining a function which takes into account also how social distancing measures and people's growing awareness of infection's risk evolves as time progresses. This also permits to reconstruct a continuous-time profile of SARS-CoV-2 reproduction number with a resolution never reached before in the literature. When applied to data collected in Lombardy, the most affected Italian region, our model illustrates how people behaviour changed during the restrictions and its importance to contain the epidemic. 
Results also indicate that, at the end of the lockdown, 
around $12\%$ of people in Lombardy and $5\%$ in Italy was affected by SARS-CoV-2.
Then, we discuss how the situation evolved after the end of the lockdown showing that the reproduction number is dangerously increasing in the last weeks due to holiday relax especially in the younger population and increased migrants arrival, reaching values larger than one on August 1, 2020. Since several countries still observe a growing epidemic, including Italy, and all could be subject to a second wave after the summer, the proposed reproduction number tracking methodology can be of great help to health care authorities to prevent another SARS-CoV-2 diffusion or to assess the impact of lockdown restrictions to contain the spread.
\end{abstract}


After its first appearance in Wuhan (China) in 2019  \cite{Fei2020,Wu2020,Guan2020}, 
SARS-CoV-2 epidemic is now affecting   
hundreds of countries over the world \cite{Velavan2020,Wittkowski2020}.
While many efforts are addressed to the development of a vaccine, 
currently the main tools to contain the pandemic 
appear social distancing measures coupled with the use of masks, massive testing and tracing approach,
or more severe restrictions like lockdown's setting \cite{Crisanti2020}.
A crucial point to increase the effectiveness 
of such actions is related to a better understanding
of COVID-19 dynamics. The ability of modeling lockdowns and 
to predict their impact on people's behaviour
is key in order to inform health-care decisions on emergency management.
This would allow to design better control strategies on  the epidemic curve, 
by gaining insight  on the number of future people who could need medical treatments.
 Modeling also allows to better assess the total number of infected, including also asymptomatic people,
and the fatality rate associated to COVID-19.\\ 

Motivated by the above arguments, mathematical modeling of SARS-CoV-2 
dynamics has been subject of intense research in the last months \cite{Bertozzii2020}.
An important class is that of \emph{compartmental models} 
where the population is assumed well-mixed and divided into categories. 
A notable example is the SIR model which includes three compartments with susceptible (S), infected (I) and removed (R) individuals \cite{Kermack1927}. 
To describe 
more complex dynamics, SIR variants can be found e.g. in 
\cite{Samanta2014,Yu2017,Funk2009,Kiss2010,Buonomo2008} where 
additional phenomena, like  
the increasing of vaccination rate, are included.
More recent extensions focus on COVID-19 pandemic and are described in
\cite{Casella2020,Lin2020,Anastassopoulou2020,Weitz2020} 
where e.g. public perception of the risk and delays effects in lockdown's setting are studied. 
An eight-compartmental model, called  SIDARTHE, has been 
also proposed in \cite{Giordano2020}.
By following some ideas introduced to describe SARS dynamics in 2004 \cite{Gumel2004},
it increases SIR complexity to discriminate between detected and undetected cases of infection.
 Another important class is the so-called {\sl spatially explicit models} \cite{Gatto2014,Riley2015}. 
They mitigate homogeneity assumptions by introducing compartments connected 
through transmission parameters to describe infection along both time and space \cite{Bertuzzo2010,Mari2019}. 
Spatial models that describe COVID-19 spread can be found, e.g., in 
\cite{OSullivan2020} and \cite{Gatto2020}
where, beyond epidemiological measurements, information
on people mobility is also exploited. 
More sophisticated models have been also proposed which include  single individual dynamics, 
i.e. the so called \emph{network models}; however their
identification is especially challenging since they contain a large number of unknown parameters 
 \cite{Keeling2005,Pastor2015,Pellis2015}.\\  
 
A common feature of all the above models is the presence of 
an important parameter which describes the virus transmission rate and takes into account also
the level of social interactions. 
We will denote it by $a(t)$, stressing its  dependence
on time $t$. To grasp its role, for sake of simplicity
a time-varying version of the SIR model can be considered.
Using $b$ to denote another parameter regulating the healing/death rate,
the differential equation which governs 
the number of infected is 
\begin{equation}\label{SIReqdiffI}
\dot{I}(t)=a(t) S(t)I(t)- b I(t).
\end{equation}
One can thus see that the contact rate $a(t)$
establishes the interaction level between  
susceptible $S(t)$ and infected $I(t)$ people, hence
regulating
the virus transmission rate. The ratio 
\begin{equation}\label{RepNumb}
\gamma(t) = \frac{a(t)}{b}
\end{equation}
defines a fundamental epidemiological variable, called \emph{reproduction number}, which 
represents the average number of infections per infected case 
and, thus, measures the disease infectivity level.\\   
Since the restrictions set 
to contain COVID-19 outbreak aim to reduce the value of the reproduction number,
one fundamental issue is to characterize in mathematical terms their impact on $a(t)$.
In this regard, compartmental or spatial models
use parametric descriptions 
of $a(t)$ by introducing an unknown 
vector of finite dimension.
In particular, e.g. as described in 
\cite{Crisanti2020,Bertozzii2020,Gatto2020}, it is common practice to adopt just two parameters
$a_1,a_2$ to quantify the two different levels of social interactions present before and after the lockdown.
Letting $t^*$ indicate the lockdown's instant,
the time-course of $a(t)$ is so given by 
\begin{eqnarray}\label{PiecewiseConstant}
\quad \qquad a(t)=  \left\{ \begin{array}{cl}  
    a_1  &  \quad  \mbox{if} \ \  t< t^* \\ 
    a_2   &   \quad \mbox{if} \ \  t\geq t^*.   
\end{array} \right.
\end{eqnarray}
An extension can be found in 
\cite{Flaxman2020}, 
where the reproduction number  is assumed piecewise constant on a finite number of 
intervals. The contact rate is allowed to change its value
only when new restraints are introduced. 
Estimation is then performed using a stochastic hierarchical model
which however requires significant Bayesian prior information.
Finally, an approach whose unique aim is to reconstruct 
the reproduction number over time (without using compartmental models) 
can be found in \cite{Wallinga2004} where estimates 
based on the observed time of symptom onset are derived.\\

\begin{figure*}
\center
{\includegraphics[scale=0.5]{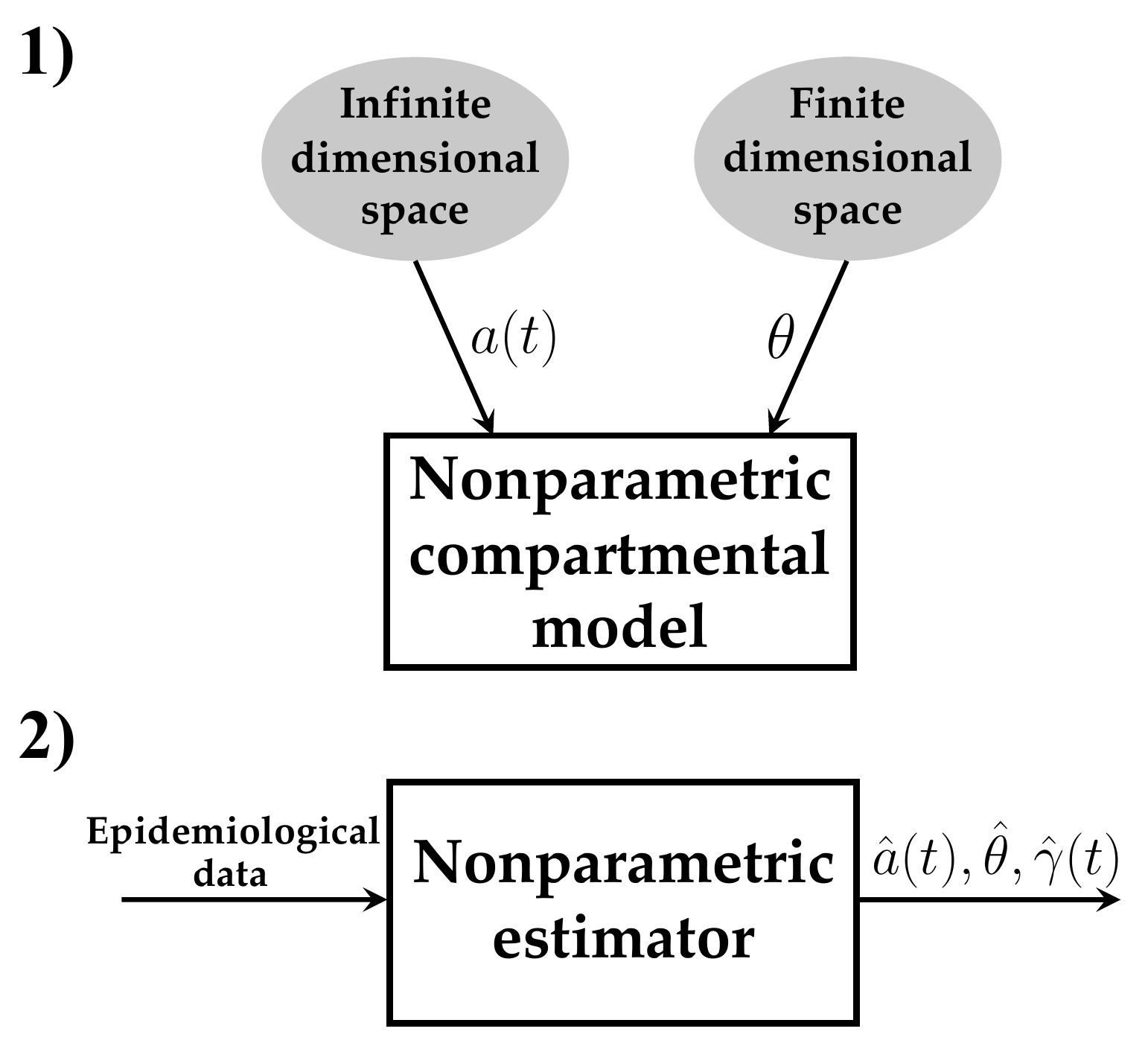}}
\caption{\emph{Top} The figure illustrates a novel class of nonparametric compartmental models.
They depend on a finite-dimensional vector $\theta$ and on a function $a(t)$ which models the virus
transmission rate and is assumed to belong
to an infinite-dimensional space. The time-course of $a(t)$ takes into account also how 
people's social interactions and awareness of infection risk evolve in time e.g. during a lockdown.
\emph{Bottom} The nonparametric estimators developed in this paper are able to map epidemiological data,
e.g. the number of diagnosed infected or hospitalized people, into the estimates of 
of the variables $a(t),\theta$ entering the nonparametric compartmental model. 
This allows also to reconstruct the time-course of the reproduction number
$\gamma(t)$ defined in (\ref{RepNumb}), which is key to monitor the epidemic evolution, 
as well as the evolution of other crucial variables like the number of infected $I(t)$.
}
\label{Fig1}
\end{figure*}

The main novelty present in this work
consists of assuming that $a(t)$ belongs to an infinite-dimensional
space containing a very rich class of functions able to approximate 
any continuous time-course. Our model is graphically depicted on the top  
of Fig. \ref{Fig1} where two different kinds of (unknown) variables are introduced.
The first one is an unknown 
finite-dimensional vector $\theta$ which can e.g. contain the parameter $b$
entering (\ref{SIReqdiffI}). The second one is the function $a(t)$
which indeed comes from an infinite-dimensional space. 
These two variables  fully define a compartmental model
whose outputs include the temporal profile of infected $I(t)$ and of removed 
$R(t)$ (who represent people who die or heal after being infected). We stress that the function $a(t)$ may complement
any kind of compartmental (or also spatial) model, the time-varying SIR being just one example.
Hence, Fig. \ref{Fig1} defines an entire novel class of \emph{nonparametric compartmental models}.
The bottom part of the same figure graphically describes a class of nonparametric estimators
able to identify such 
models using epidemiological data like e.g. 
the number of diagnosed infected or hospitalized. \\

As a case study to show the potential of this novel non-parametric modeling identification strategy we consider the Italian scenario, and in particular the Lombardy region which has been the most affected (see also Results).
Many people in Italy underwent screening for COVID-19 
starting from the end of February 2020.
Data are 
publicly available and are collected on a daily basis  \cite{Data2020}.
Some of them are displayed in Fig. \ref{Fig1abc} where one can see: 1) number of people diagnosed as infected by COVID-19;
2) number of infected and hospitalized; 3) number of infected people who are in intensive care.
However, the data on the number of people diagnosed as infected by COVID-19 (point 1) have important limitations. They do not give an accurate information on how many subjects were infected exactly at a certain day due to delays in the swabs processing. In addition, the amount of performed swabs and the criteria used to select people who are tested may vary in time. In contrast, data describing the number of hospitalized people  (point 2) and, even more, patients in critical care (point 3) appear more reliable and informative. Hence, to identify the model, our estimator will exploit the number of people in critical care. In particular, an assumption which appears statistically reasonable is to assume that the number of infected is proportional, through a multiplier $H$, to the number of people in intensive care. Hence, the scalar $H$ represents another unknown component of the parameter vector $\theta$.\\

However, even when a relatively simple time-varying SIR model is adopted, the resulting estimation problem 
turns out to be ill-posed \cite{Bertero1}.
In fact, beyond $\theta$,  the infinite-dimensional function $a(t)$ 
must be inferred from a finite number of measurements. 
Despite these difficulties, in Methods we show that it is possible to design an estimator which
is purely data-driven and does not need additional epidemiological or clinical
prior information. The problem is set in the framework of 
regularization in stable reproducing kernel Hilbert spaces (RKHSs) 
whose importance in machine learning and system identification problems has been recently described in
\cite{PillonettoDCNL:14,PillonettoMF2020}. 
Here we just recall that one fundamental peculiarity
of these spaces is that their complexity is regulated by a stability parameter that also needs  
to be estimated from data. In our setting, it regulates how fast $a(t)$, and 
the reproduction number $\gamma(t)$,
is expected to decrease after the lockdown. As described in Methods, implementation of the resulting estimator,
which is graphically depicted in the bottom part of Fig. \ref{Fig1}, is far from trivial since
both $\theta$ and $a(t)$ are nonlinearly related to the intensive care data.
Nevertheless, an efficient optimization scheme providing the desired estimates can be defined,
together with a Markov chain Monte Carlo scheme \cite{Gilks}
which returns confidence intervals around them.

\section*{Results}

\begin{figure}
	\begin{center}
		\begin{tabular}{c}
			\hspace{-.2in}
			{ \includegraphics[scale=0.23]{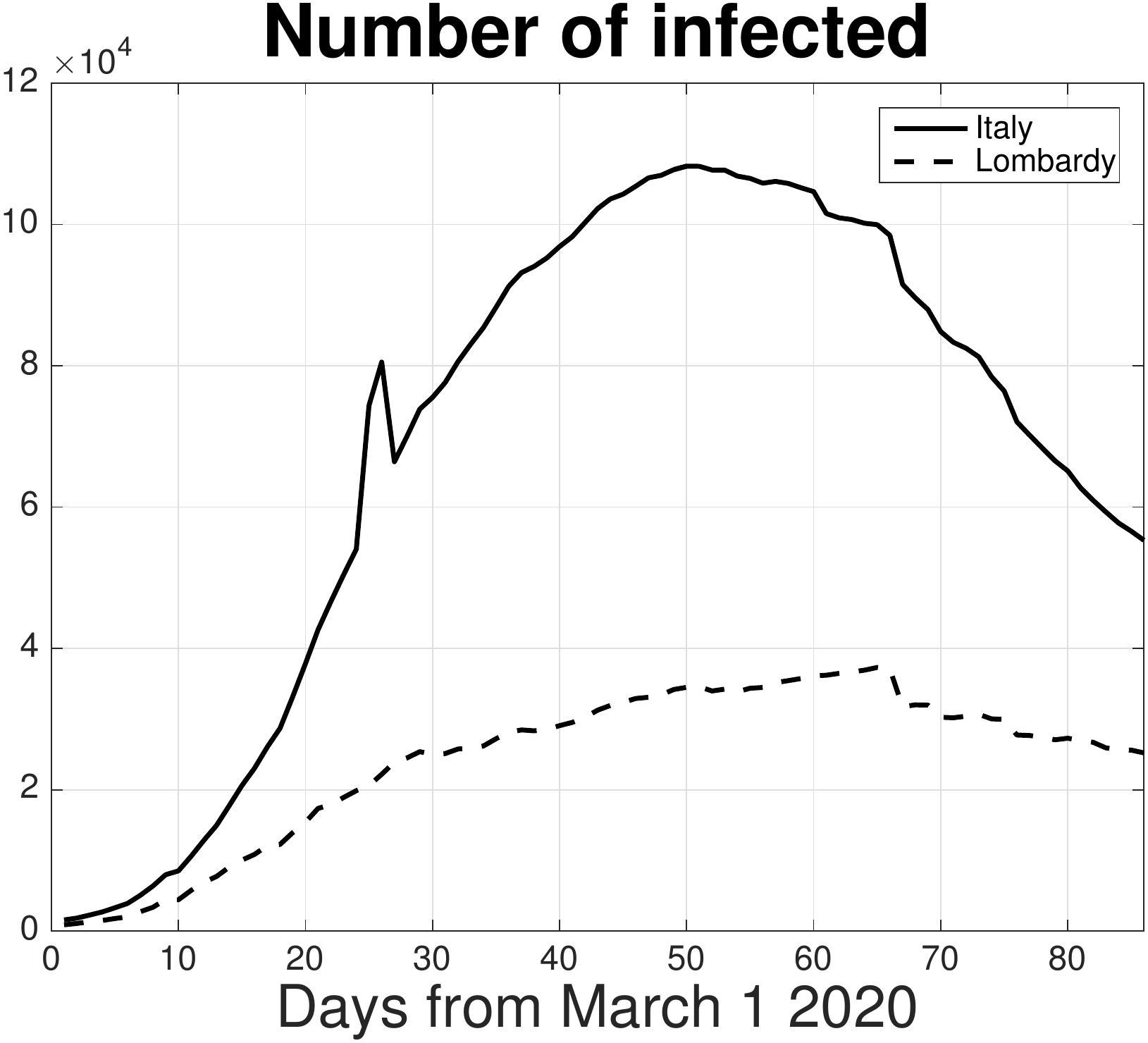}} \
			{ \includegraphics[scale=0.23]{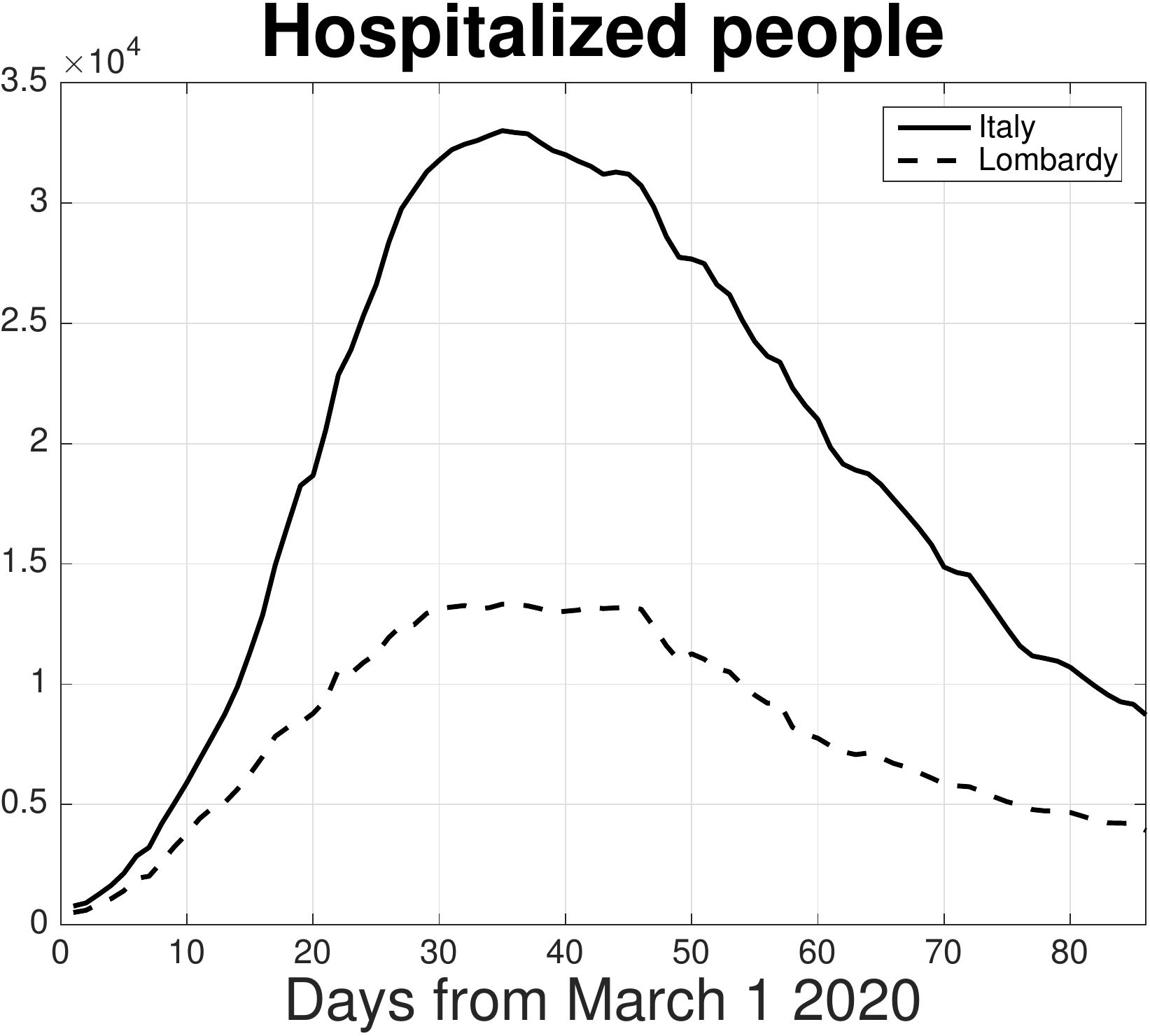}} \
			{ \includegraphics[scale=0.23]{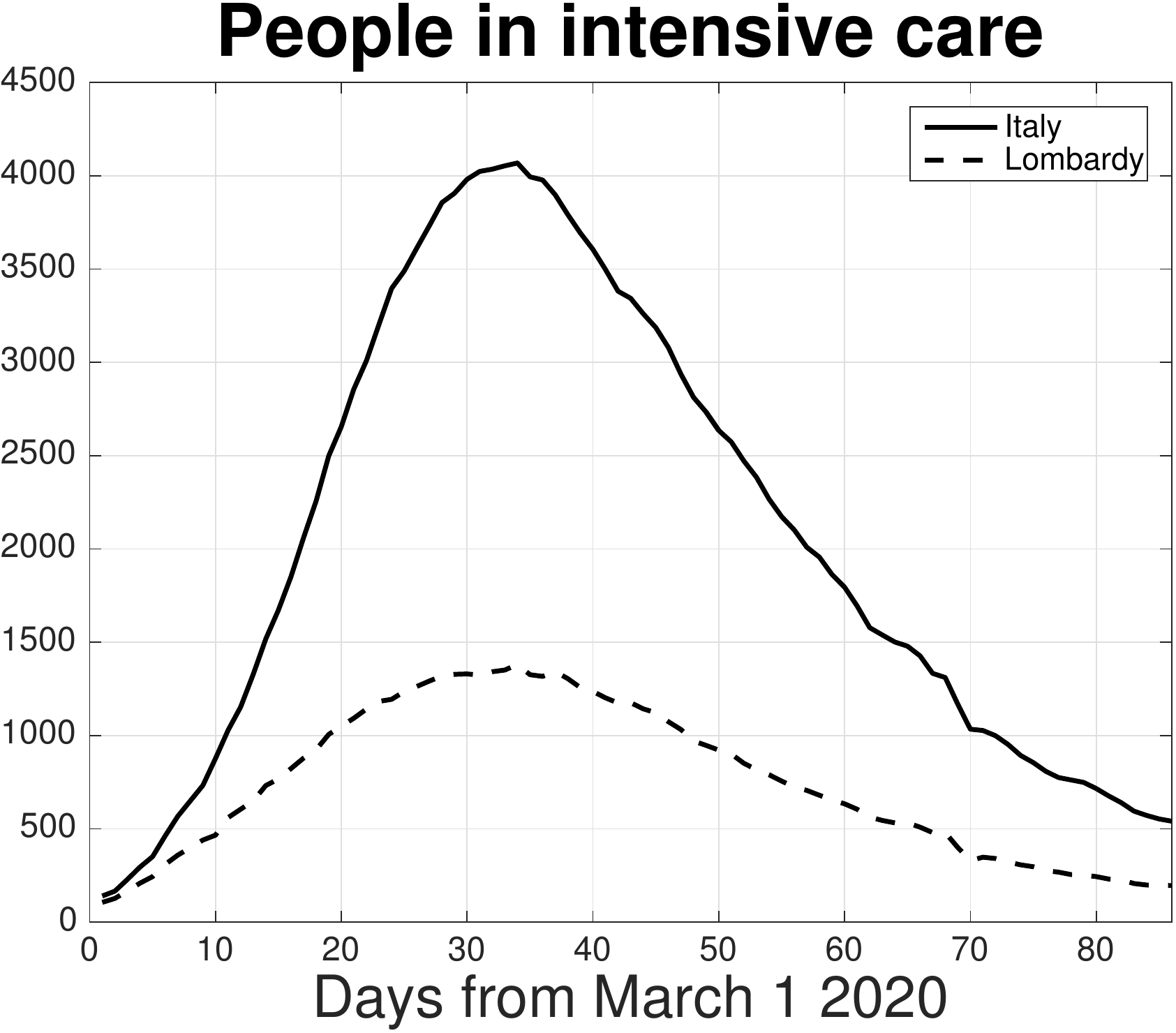}} 
		\end{tabular}
		\caption{Diagnosed infected (left), hospitalized people (middle) and people in critical care (right)
		in Italy (solid line) and Lombardy (dashed). Instant 1 corresponds to March 1,  2020. 
		One can see that, in the last months, Lombardy has collected almost $40\%$ of infected and hospitalized Italian people.
		This makes this region an important case study to apply the novel non-parametric modeling identification strategy here proposed.
		} \label{Fig1abc}
	\end{center}
\end{figure}

We will now describe results obtained by our nonparametric model identification method,
with dynamics of infected people described by (\ref{SIReqdiffI}), 
estimated by using intensive care data collected in Lombardy.
This  region contains around ten million people and is a natural candidate to tune our model since most of the Italian outbreak happened there.
In the last months it has collected almost $40\%$ of infected and hospitalized Italian people  and around
17000 people died in Lombardy due to COVID-19 as of July, 2020, see Fig. \ref{Fig1abc}.
The large diffusion is also revealed by some small preliminary studies
on antibody responses to virus performed on blood donors.
Authors of the work \cite{Valenti2020} considered 
a random sample of 789 blood donors in Milan.
At the start of the outbreak, on February 24,
the overall seroprevalence of SARS-CoV-2 was $4.6\%$ $(2.3\%-7.9\%)$.
In the town Castiglione d'Adda, the epicentre of the outbreak,
at the beginning of April $70\%$ out of 60 
asymptomatic blood donors had the antibodies \cite{LaStampa2020}.
Outcomes from Lombardy can then be used to achieve estimates 
of the number of infected at the Italian level by exploiting the assumption on the multiplier $H$.\\

To correctly interpret the following results, it is worth recalling that 
Italy has been the first country in Europe to set nationwide restrictions by 
introducing the lockdown to the whole territory on March 9, 2020.
Restrictions have then been first further reinforced 
and then gradually relaxed. Almost all the activities re-opened
on May 18, 2020. For this reason, we will first  exploit data on the temporal window
going from March 1 to May 17, 2020 to study how the lockdown has affected the contact rate
and the reproduction number. Data collected starting from May 18, 2020 will be then used to
study how the situation evolved after the end of the lockdown.\\

\begin{figure*}
\center
{\includegraphics[scale=0.5]{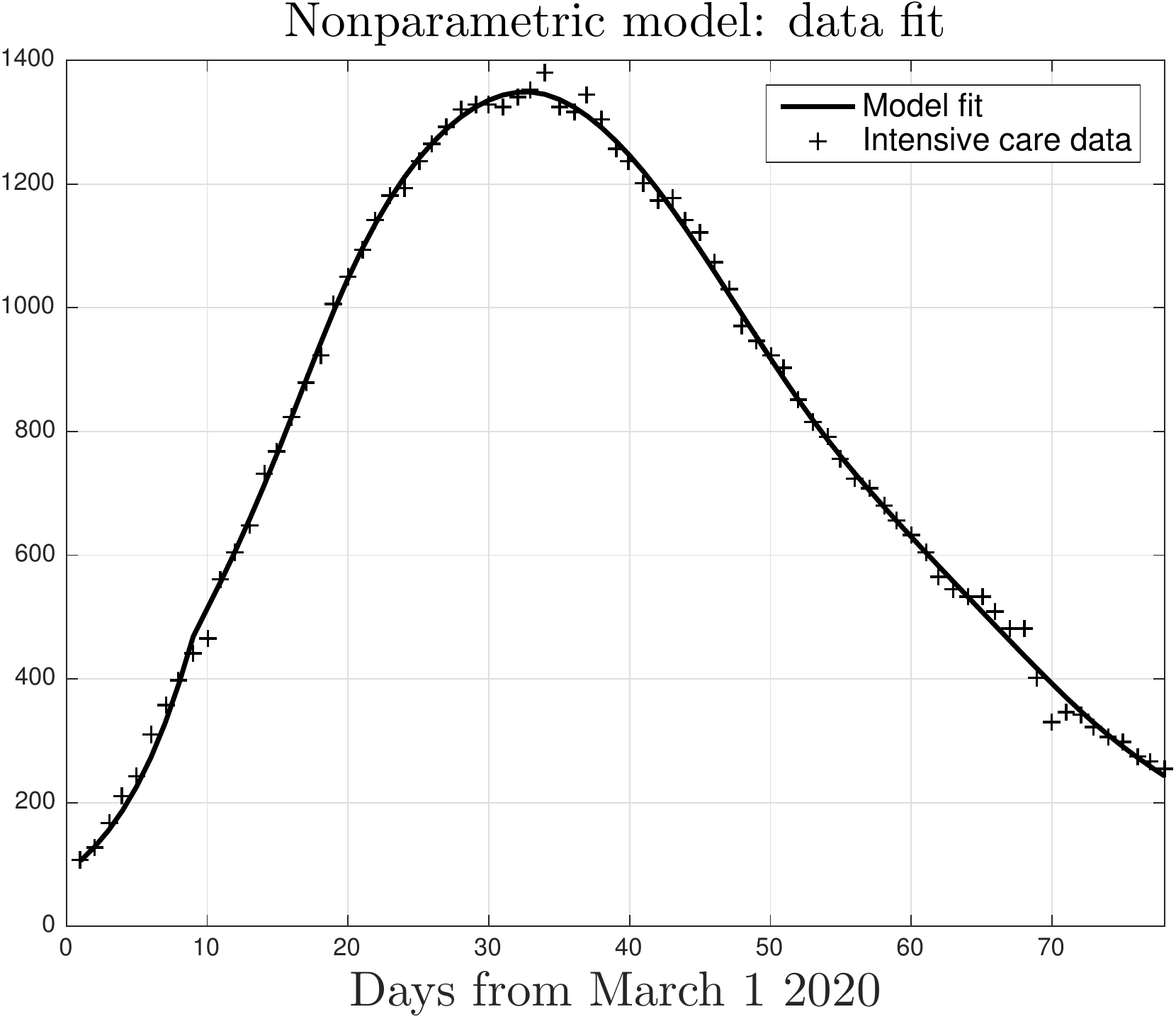}}
\caption{Fit of the intensive care data collected in Lombardy obtained by the nonparametric compartmental model.
The interaction between infected, $I(t)$, and susceptible, $S(t)$, people is described through (\ref{SIReqdiffI})
with $a(t)$ assumed to belong to an infinite-dimensional space.}
\label{Fig2a}
\end{figure*}

\begin{figure*}
\center
{\includegraphics[scale=0.5]{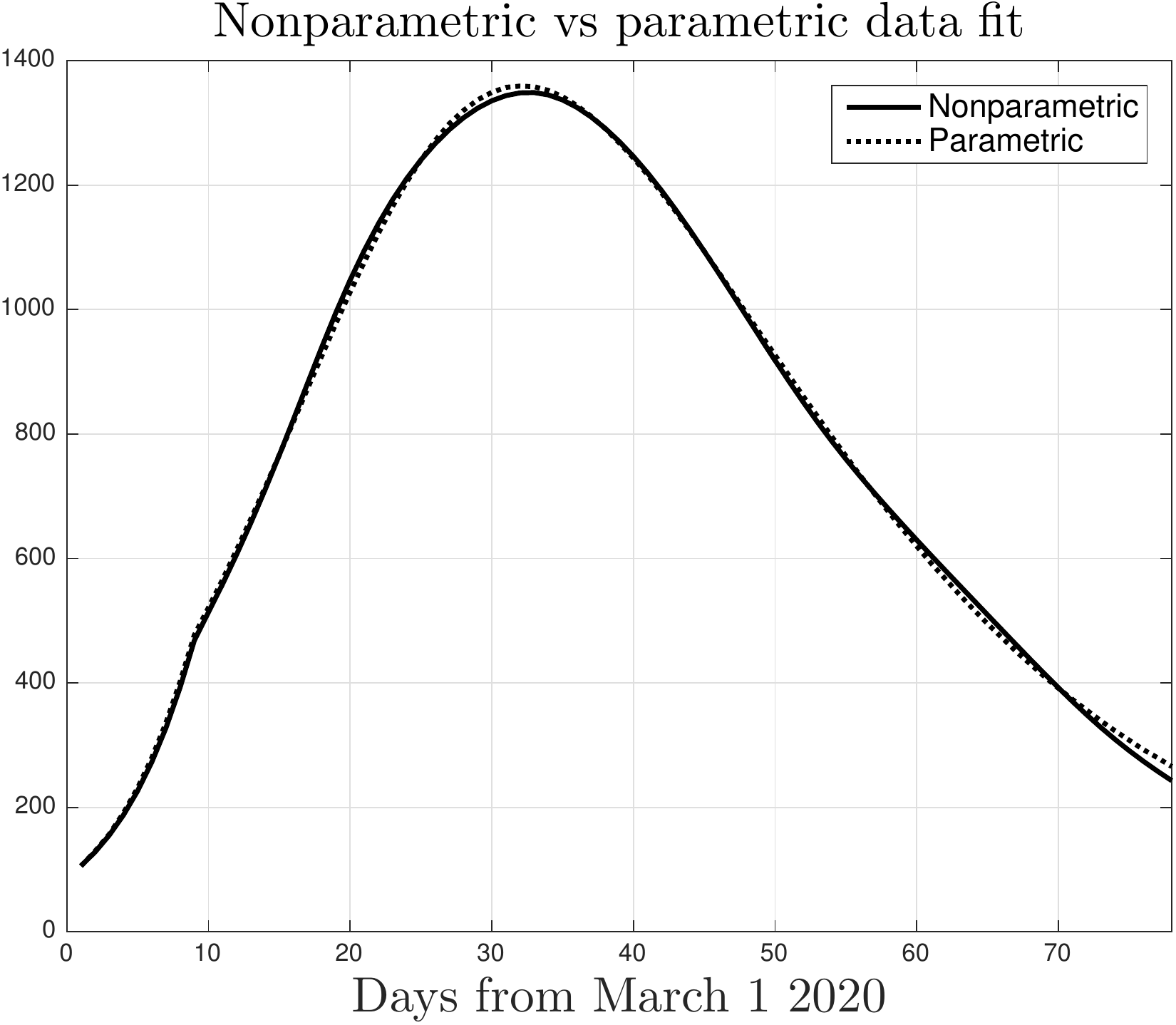}}
\caption{Nonparametric (solid line) and parametric (dotted) fit of the intensive care data collected in Lombardy.}
\label{Fig2b}
\end{figure*}

Fig. \ref{Fig2a} reports the intensive care data collected in Lombardy together with  
the fit returned by our nonparametric technique. One can see that 
the model is able to well describe the observational data.
Fig. \ref{Fig2b} compares the data fit obtained by the nonparametric (solid line) and the parametric model (dotted),
this latter using (\ref{PiecewiseConstant}) to describe $a(t)$.
Apparently, there is no real difference 
in the data fit but we will show the very different results obtained with the nonparametric technique.\\

\begin{figure*}
\center
{\includegraphics[scale=0.5]{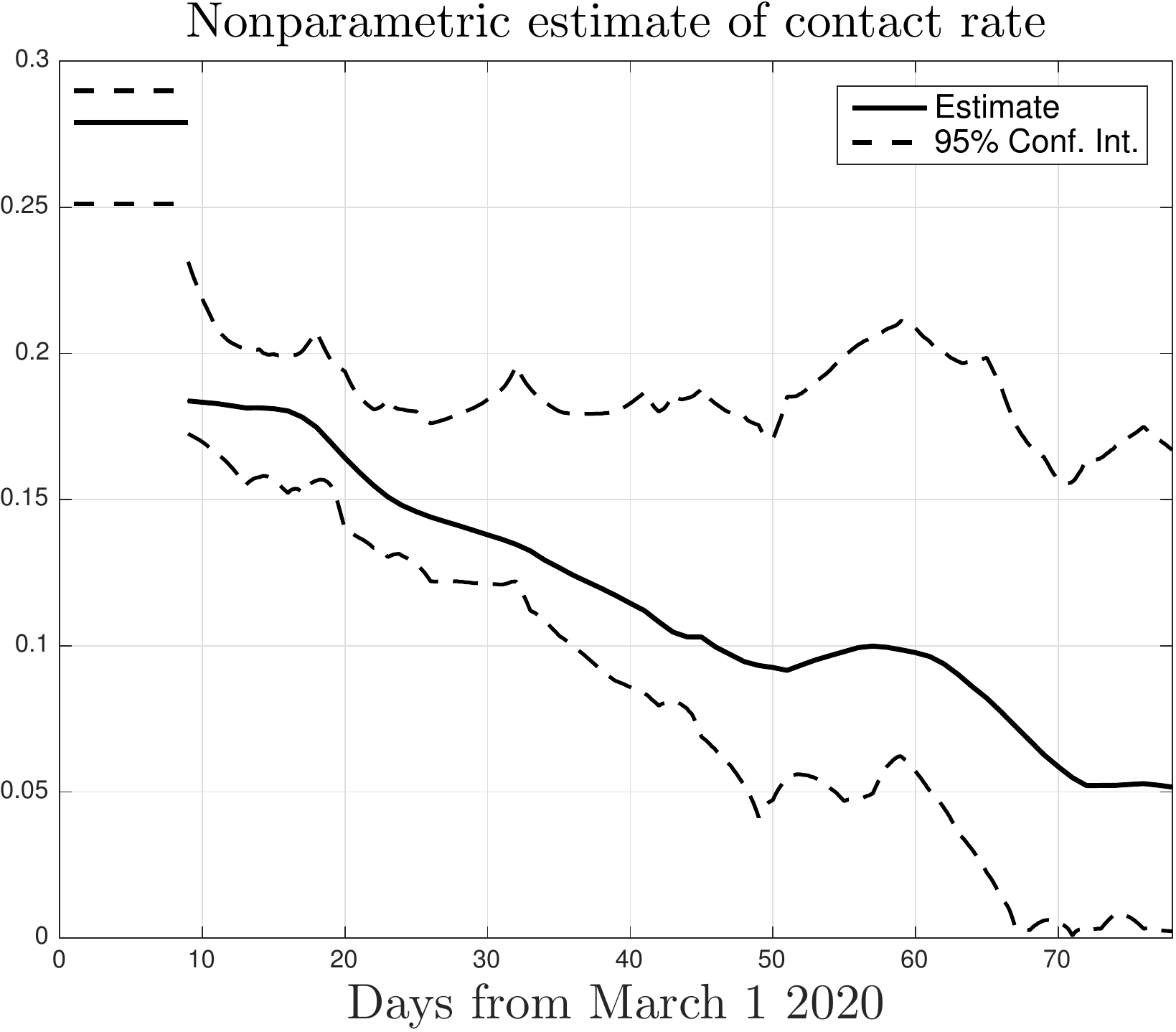}}
\caption{Nonparametric estimate of the function
$a(t)$ entering (\ref{SIReqdiffI}) together with $95 \%$ confidence intervals. Its time-course describes how 
social interactions
and, hence, the virus transmission rate, 
changed in Lombardy before and after the lockdown which started on March 9, 2020.}
\label{Fig3}
\end{figure*}

Fig. \ref{Fig3} plots the estimate of
$a(t)$ (solid line) which 
describes how the level of social interactions in Lombardy evolved
in time. The same figure also reports $95 \%$ confidence intervals (dashed).
Uncertainty bounds are quite asymmetric and not so small,
pointing out the difficulty of the problem,
 but results appear somewhat significant.
One can in fact see that, after March 9 (the beginning of the lockdown),
$a(t)$ decreases from 0.28 to 0.184. 
Such value in practice remains constant for almost ten days.
After March 20, the consequences of the restrictions become
more pronounced and the curve decreases until
April 20 (day 50 on the $x$-axis). 
The contact rate then diminishes again starting 
from the beginning of May (61 on the $x$-axis),
reaching a plateau value equal to 0.05 ten days before the lockdown's end.
Overall, the reconstructed profile of $a(t)$ seems really realistic,
also in view of the fact that
restrictions were further strengthened after the first lockdown. 
Beyond delays in observing the restrictions effects,
our estimated trend surely incorporates,
with a resolution never reached before in the literature,  
also other complex phenomena. Examples regard increments in
people perception
of infection risk and in the use of precautions, like protective masks,  
which greatly helped to control the virus spread.\\

\begin{figure*}
\center
{\includegraphics[scale=0.5]{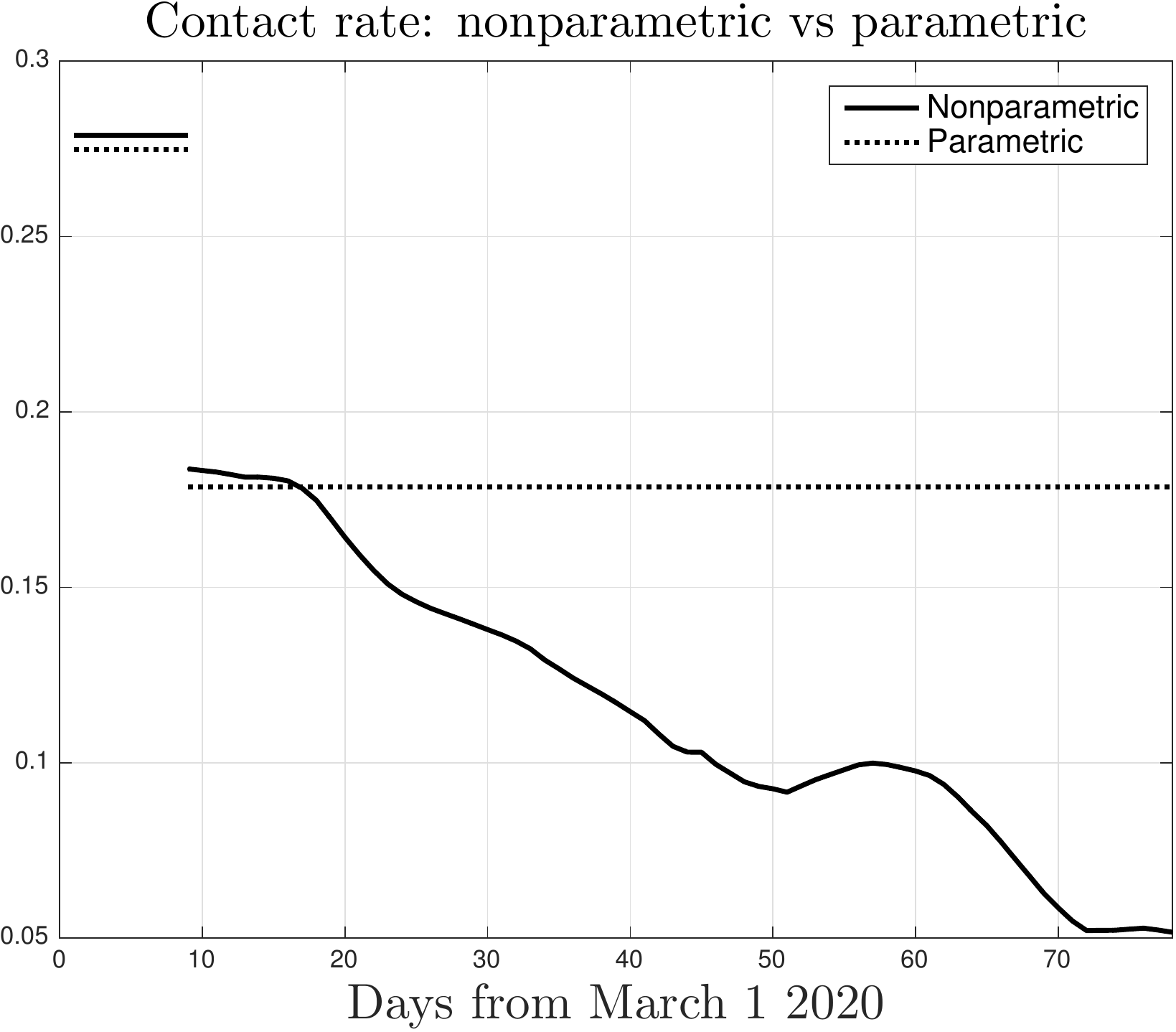}}
\caption{Nonparametric (solid line) and parametric (dotted) estimate of the function $a(t)$ describing 
how the level of social interactions change in time. The nonparametric estimator assumes that $a(t)$
belongs to an infinite.dimensional space whereas the parametric approach models it
as piecewise constant according to (\ref{PiecewiseConstant}).}
\label{Fig4}
\end{figure*}

The importance of our new technique is also illustrated in Fig. \ref{Fig4} 
which compares the nonparametric estimate of $a(t)$ (solid line, the same displayed in Fig. \ref{Fig3})
and the parametric one (dashed) where $a(t)$ can assume only two values according to (\ref{PiecewiseConstant}).
Obviously, the parametric reconstruction is unable to track the time-varying  impact of the restrictions in Lombardy. 
It is interesting to see that, for all the duration of the lockdown, the 
estimate of the social interactions level is around 0.18, close to that returned by the nonparametric estimator at the beginning of the lockdown. So, all the considerations regarding the evolution of people behaviour are lost using the classical approach. 
This inevitably leads to overestimation of SARS-CoV-2 transmission rate.\\

\begin{figure*}
\center
{\includegraphics[scale=0.5]{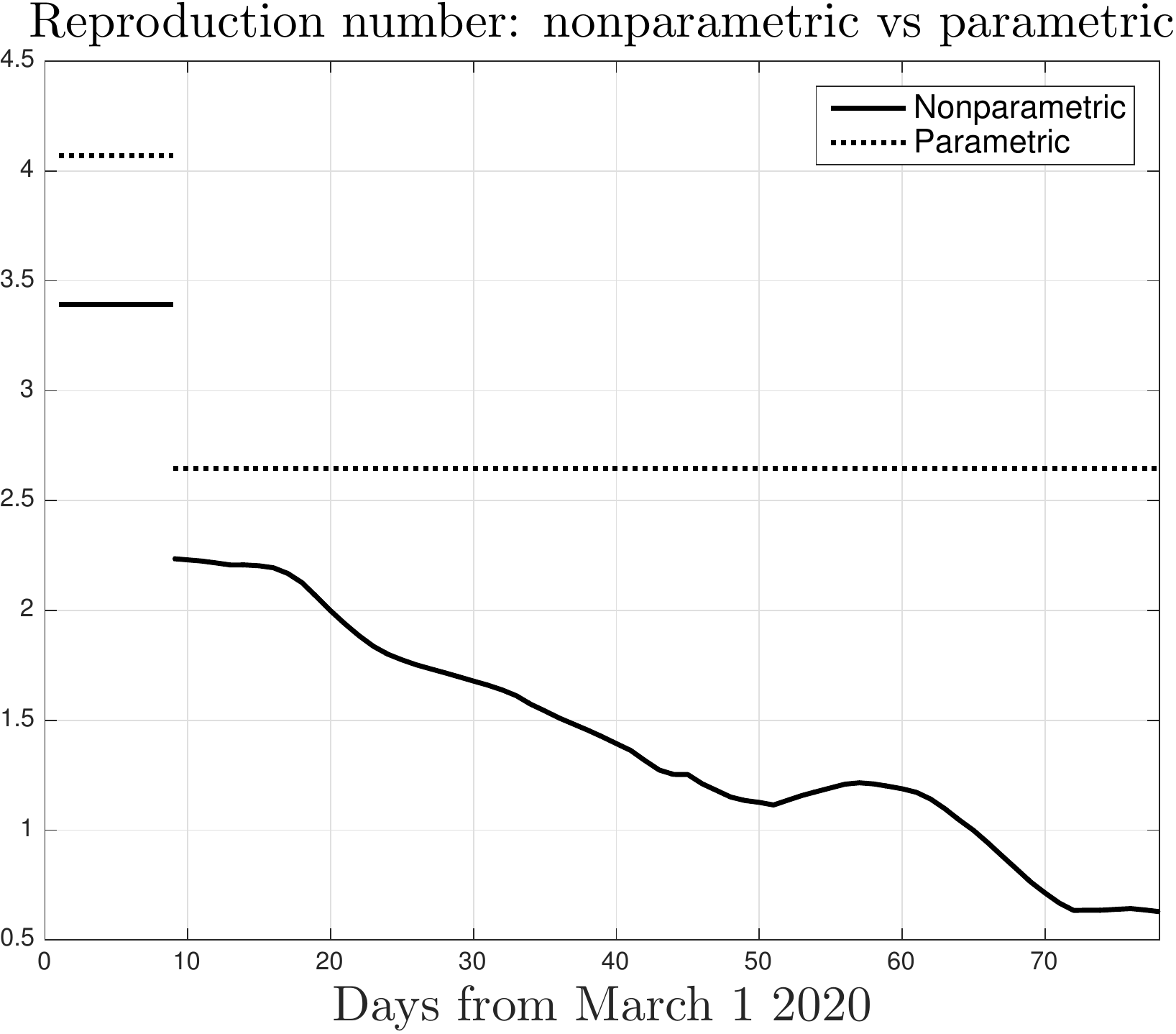}}
\caption{Nonparametric (solid line) and parametric (dotted) estimate of the reproduction 
number $\gamma(t)$ in Lombardy, as defined in (\ref{RepNumb}).}
\label{Fig5}
\end{figure*}

\begin{figure*}
\center
{\includegraphics[scale=0.5]{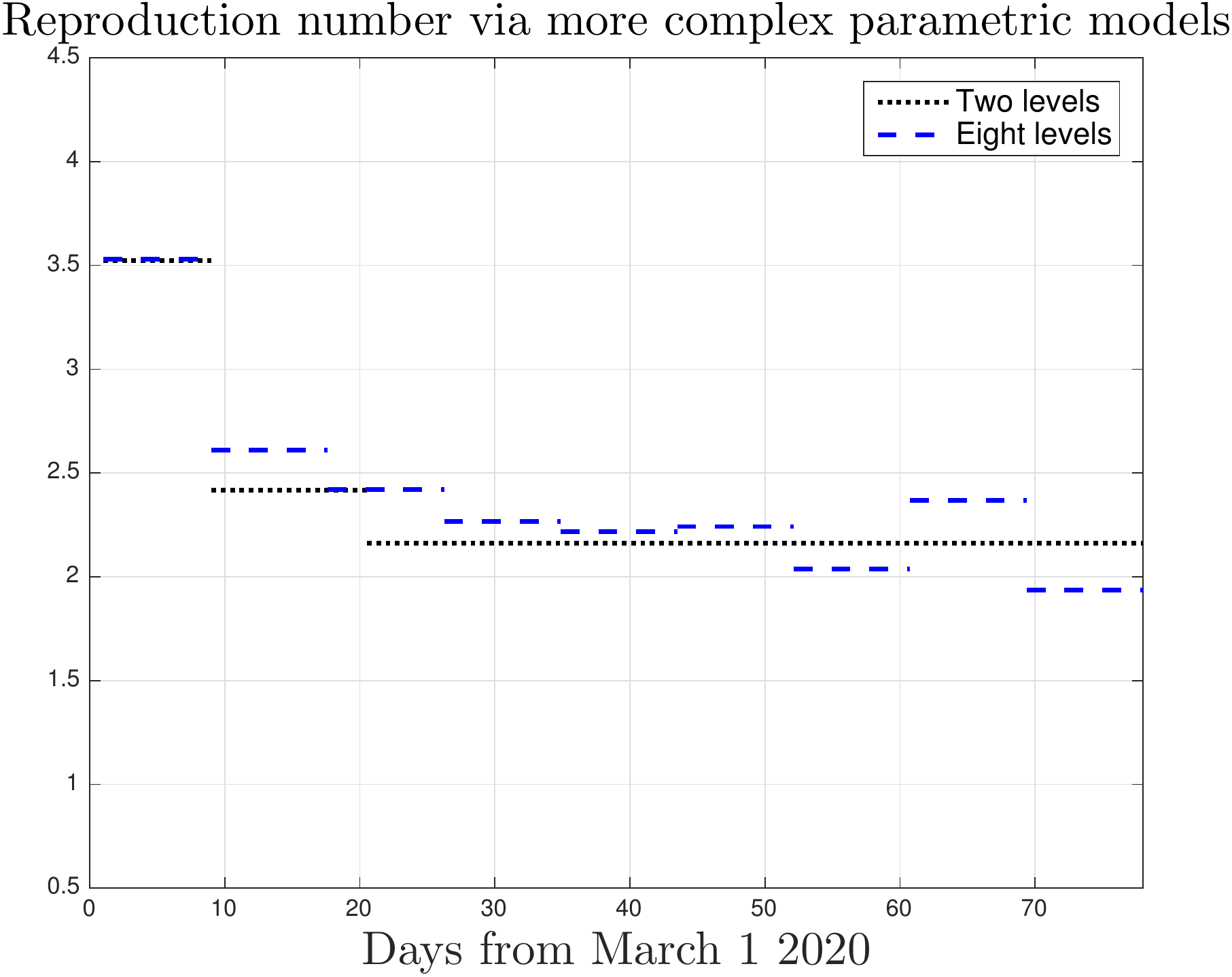}}
\caption{Parametric estimate of the reproduction 
number  in Lombardy with the contact rate $a(t)$ allowed to assume
two (dotted) or eight (dashed) different values during the lockdown.}
\label{Fig5b}
\end{figure*}

Fig. \ref{Fig5} displays the nonparametric estimate  of the reproduction 
number in Lombardy (solid line) as defined in (\ref{RepNumb}). 
Tracking $\gamma(t)$ is crucial 
to understand if the restraints are effective: the epidemic is under 
control when it is smaller than one.
Our results suggest that, before the lockdown, 
its value was somewhat large,
 around 3.4. After March 9 
 it decreased to 2.3 and values close to one
 were obtained only after 40 days from
 the beginning of the lockdown (day 50 on the $x$-axis).
 The reproduction number reached its minimum value, around
 0.6, just a few days before the end of the lockdown.\\
 The same figure also reports the parametric estimate  of the reproduction 
number in Lombardy with $a(t)$ constrained to assume only two values (dotted line).
Differently from the nonparametric technique, this approach 
overestimates the reproduction number and
does not allow to understand when the epidemic is under control.
Failure of the parametric method has been
assessed 
also using more complex parametrizations.
In particular, Fig. \ref{Fig5b} also reports estimates from two parametric 
models with the contact rate $a(t)$ that may assume two values (dotted),
with switching instant determined from data,
or eight (dashed), in this case over uniformly spaced intervals.
In both the cases the information obtained regarding the 
reproduction number is unsatisfactory and its value remains larger than one 
during all the lockdown.\\



\begin{figure*}
\center
{\includegraphics[scale=0.5]{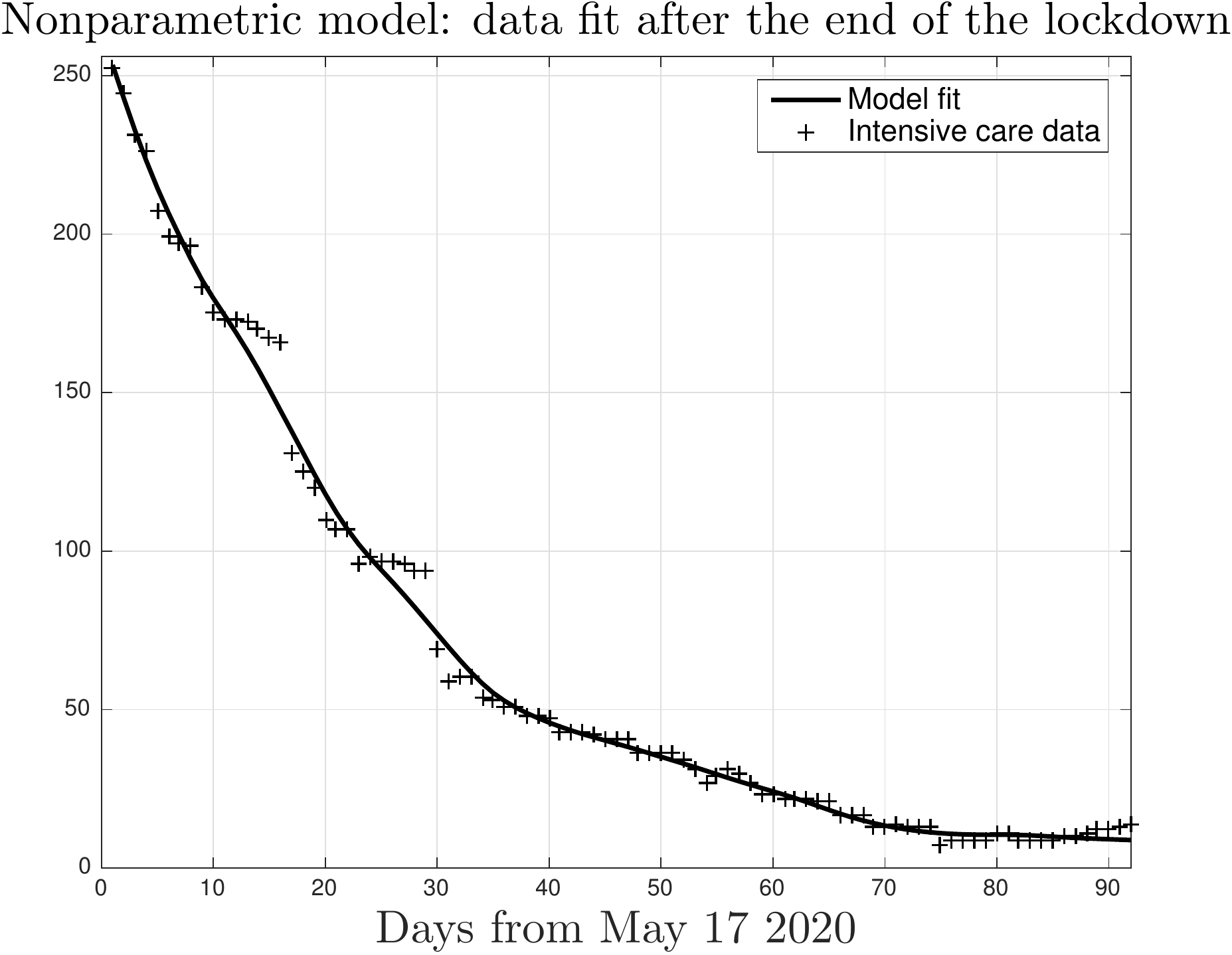}} \vspace{0.6cm} \\  {\includegraphics[scale=0.5]{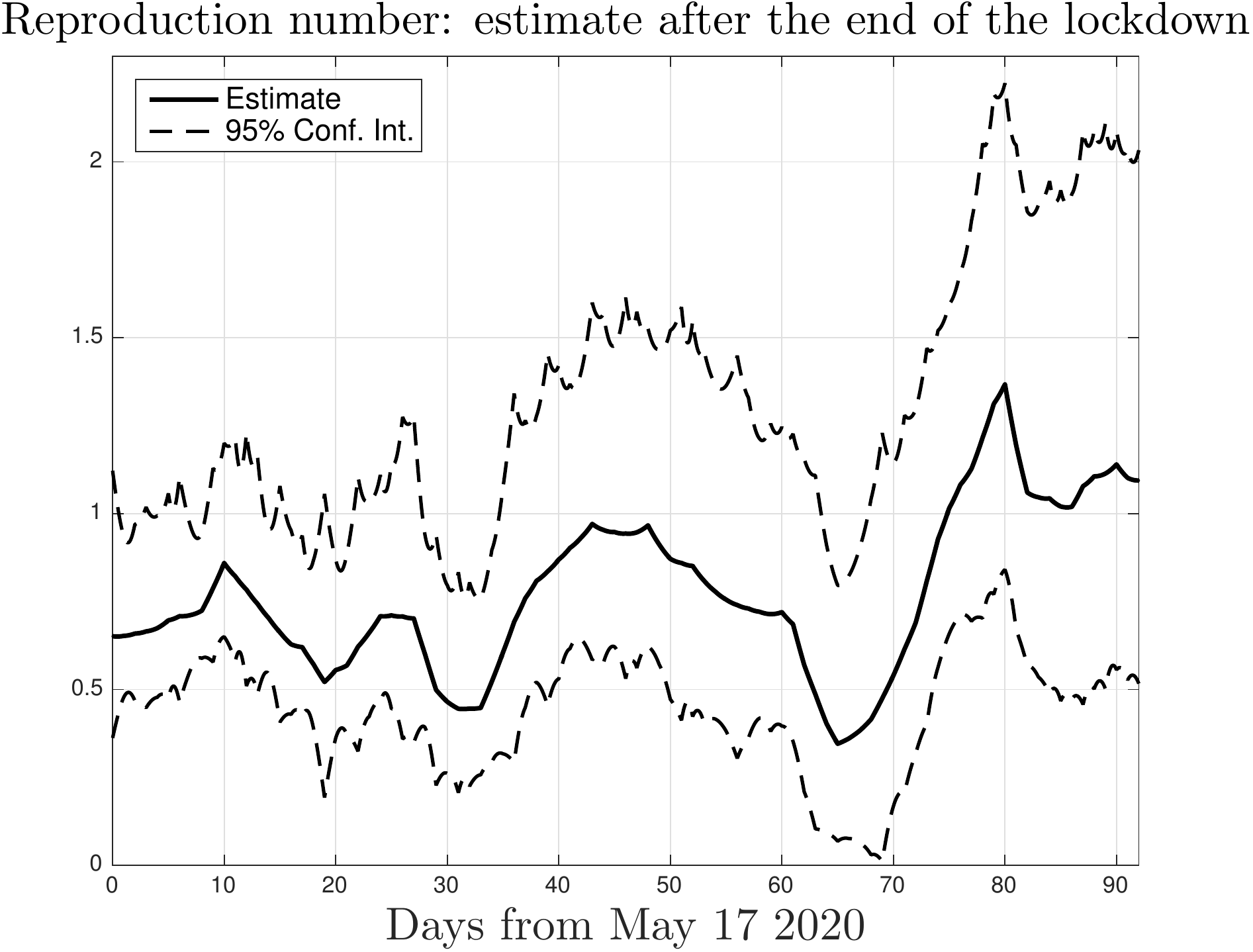}} 
\caption{ \emph{Top} Nonparametric model fit of the intensive care data observed in Lombardy after the end
of the lockdown. \emph{Bottom} Nonparametric estimate of the reproduction number in Lombardy after the end of the lockdown together with $95\%$ confidence intervals (dashed).}
\label{Fig9}
\end{figure*}
 
After documenting the spread evolution in Lombardy during the lockdown,
 the nonparametric model is now exploited to understand how the situation evolved in the last months.
For this aim, we use the intensive care data collected from May to August, 2020. They are shown in the top panel
of Fig. \ref{Fig9} together with the nonparametric model fit. The bottom panel then reports the 
estimated time-course of the reproduction
number  in Lombardy after the end of the lockdown. On May, and for the most part of June, the 
estimate is below the critical threshold. But at the end of June (around day 40 on the $x$-axis) and in the first part of July 
the reproduction number was close to one. Then, after decreasing (starting from day 60 on the $x$-axis),
it increased becoming larger than one on the first of August.
The maximum value 1.36 was reached on August 5 while 
on August 17, the last day here considered, the reproduction number was close to 1.1.
Our outcomes suggest that the adoption of social 
distancing measures and protections like masks greatly decreased during the summer season.
Policy makers need to carefully consider this to avoid a second wave of SARS-CoV-2 spread in Italy during 
the next months.\\

 \begin{figure*}
\center
{\includegraphics[scale=0.5]{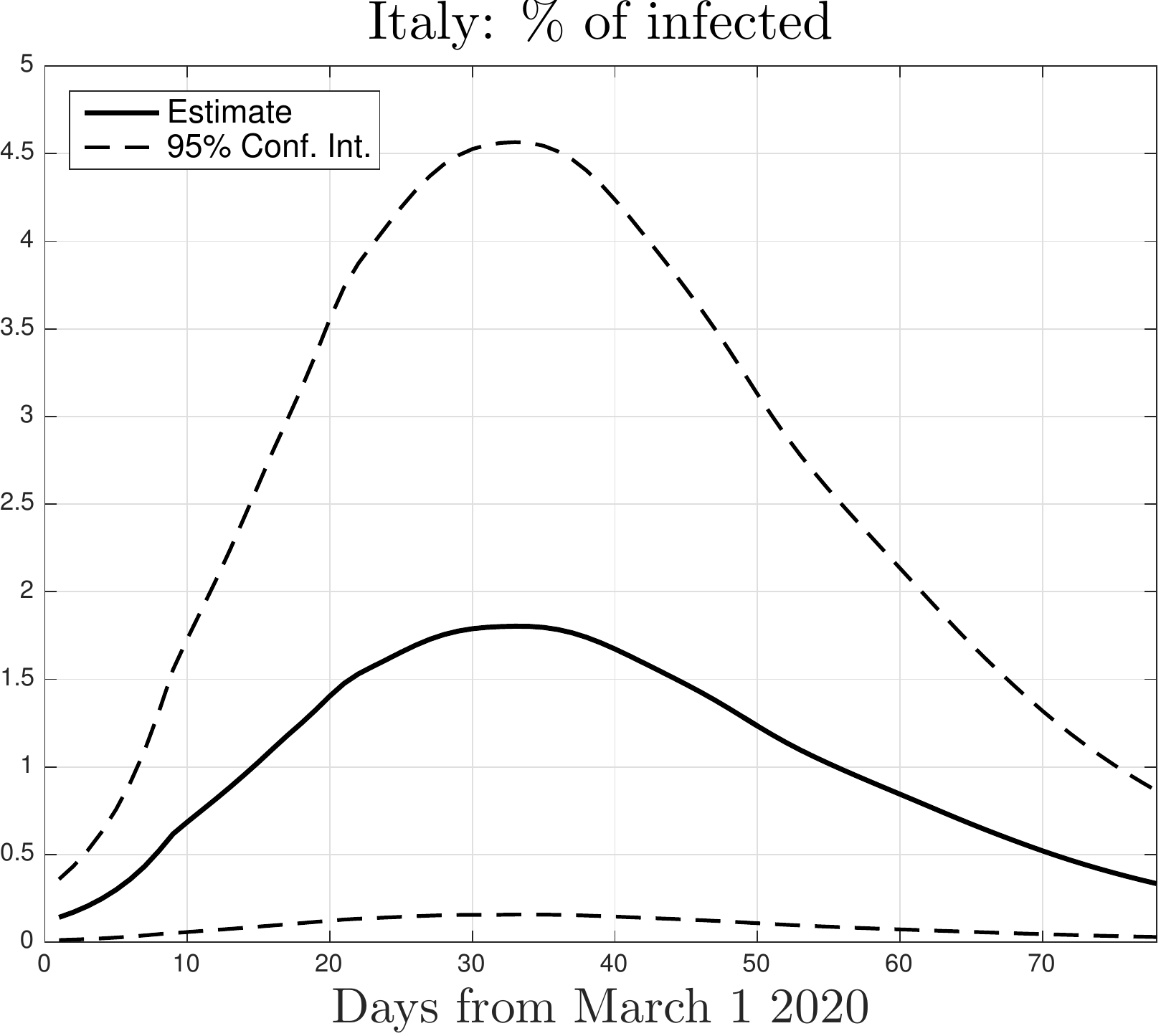}} \vspace{0.5cm} \\  {\includegraphics[scale=0.5]{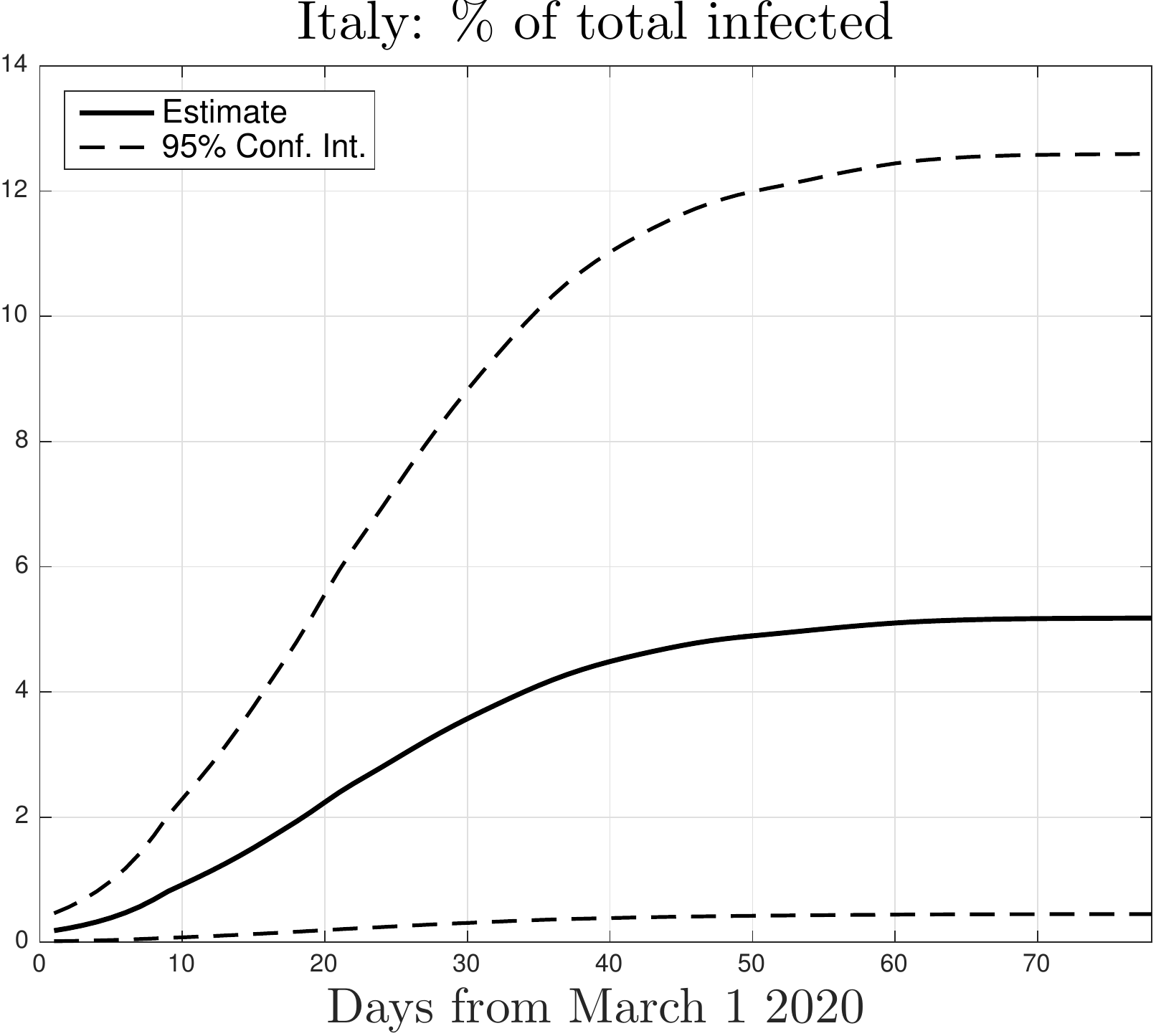}} 
\caption{Estimated number of infected $I(t)$ (top) and of 
total infected $I(t)+R(t)$ (bottom) in Italy, together with $95\%$ confidence intervals (dashed). }
\label{Fig6}
\end{figure*}

The nonparametric approach can also infer  the number of infected in Italy. A very recent study 
has detected the presence of antibodies against SARS-CoV-2 in $2.5\%$ of Italian population and $7.5\%$ in Lombardy, e.g. see \cite{Repubblica2020}. These estimates are however affected by two sources of uncertainty. The first one is due to the size of the sample (only around 60 thousand out of 60 million people underwent the test in Italy). 
 The second one is the fact that the recent Nature Medicine report \cite{Long2020} shows that antibody levels can drop significantly during recovery. The levels could become undetectable within 2-3 months, especially in asymptomatic or people who showed mild symptoms.
To account for these uncertainties, still considering Lombardy as case study, we interpret the percentage of infected as a nonnegative random variable with mean 7.5. Among the infinite types of probability distributions compatible with such information we then choose that maximizing the entropy \cite{Jaynes,Papoulis}. It turns out that the least committing prior on the percentage of infected deriving from the antibody tests is an exponential distribution whose mean (and SD) is 7.5.
Interestingly, when coupled with such a prior, the parametric approach 
is unable to describe the intensive care data in an acceptable way.
This means that model (\ref{PiecewiseConstant})  
is not sufficiently flexible  to trade off measurements and prior information coming from antibody tests.
The nonparametric approach is instead much more versatile 
and can well describe the observational data obtaining a fit similar to that displayed in Fig. \ref{Fig2a}. 
Our model then predicts that the estimated percentage of infected in Lombardy 
at the end of the lockdown was close to $12.5\%$. 
By projecting such result at national level, using the assumption on the multiplier $H$,
Fig. \ref{Fig6} reports the estimated
 time-courses of infected  $I(t)$ (top panel) and of 
total infected $I(t)+R(t)$ (bottom) in Italy.
The top panel shows that
the estimated value of the peak of infected was around $1.8\%$. 
 The bottom panel shows that our model 
 then predicts that almost $5.1\%$ of Italian population had been infected at the end of the lockdown.
The upper bound for the $95\%$ CI is around $13\%$, suggesting that
antibody tests could underestimate significantly 
the number of total infected.\\

 \section*{Discussion}
  
The compartmental and spatial models of SARS-CoV-2 dynamic so far adopted in the literature use parametric descriptions of the contact rate and the reproduction number. This paper shows that these approaches may lead to a wrong assessment on ongoing pandemic since they have difficulties to properly capture lockdown's time-varying impact on people behaviour. This may lead to an overestimation of the reproduction number which does not allow to well understand if and when the epidemic is under control.
Motivated by these difficulties, in this work we have developed new regularized machine learning techniques which lead to the definition of an entire new class of nonparametric compartmental models. The contact rate is not confined to live in a finite-dimensional space and the temporal profile of the reproduction number is estimated from intensive care data within a very rich family of functions.
Our approach, applied to data collected in Italy, shows that the reproduction number can embed many factors really hard to be captured by parametric models. They include significant delays in lockdowns effects since social distancing measures and use of protections like masks can greatly vary over time. The nonparametric estimator is instead able to track these phenomena with a resolution never reached before. This may greatly help in quantifying their impact to contain SARS-CoV-2 spread, allowing to convey useful messages to political decision makers.\\

We have used the Lombardy data, the most affected Italian region, as case study to show the potential of the reproduction number tracking methodology to document how people behaviour changed during the restrictions and its importance to contain the epidemic.
We have also illustrated how the situation changed after the end of the lockdown. Obtained results suggest that the adoption of social 
distancing measures and protections much decreased  during the summer 
season due to holiday relax especially in the younger population and increased migrants arrival.
This appears especially dangerous since such people behaviour could trigger a second wave of SARS-CoV-2 spread in Italy during 
the next months.
Results have been also properly extrapolated to obtain a new nonparametric estimate of the number of infected in Italy.
Even if care has to be taken in their interpretation, e.g. in view of the homogeneity assumptions underlying the
time-varying SIR model here adopted (the population has to be well mixed), results appear important and describe a 
level of epidemic diffusion in Lombardy and Italy around $12\%$ and $5\%$, respectively.\\

There is no precise appreciation of virus circulation in Lombardy, nor in Italy, from which deriving a real measure of morbidity and lethality. This latter parameter, obtained from total deaths over confirmed positive cases (35,437 over 259,345 as of August 24, 2020) actually accounts for 13.66$\%$, one of the highest estimates in the world. The cause for this exceptional event certainly stems from the extremely severe COVID-19 cumulative incidence at the pandemic onset in Lombardy and in some northern regions.  All available intensive care beds were in fact overloaded in Lombardy as well as first aid facilities, a condition that allowed a large and rapid spreading of a considerably high contagion both in the nosocomial and in the community settings.
 To have a more precise assessment of  COVID-19 lethality,
 only a few studies of seroprevalence have been then conducted in Italy, see e.g. \cite{Plebani2020}. They were performed independently by some regions or single hospital institutions by means of either accurate CLIA or ELISA blood tests or quick lateral chromatography assays, whose performance is rather questionable \cite{Krammer2020,JH2020}. The only nation-based study, a CLIA assay measuring anti-N antibodies failed to reach its designed target of 150.000 subjects stratified per age, risk, sex and location. This leads to an estimated fatality rate in Italy around $2.5\%$.
 If one considers that specific antibody testing is not always available for new pathogens and that serology can underestimate the total number of SARS-CoV-2 infected individuals \cite{Long2020}, our proposed non parametric compartmental model brings about a very valuable method to measure the real clinical and public health impact of the pandemic and its evolution. Interestingly, it provides
 an estimate of the fatality rate around $1\%$ which is now in line with
those concerning other countries and recently reported  in the literature  \cite{NatureNews2020}.\\
 
In conclusion, a number of mathematical models have been proposed to predict the biological and medical implications of the ongoing COVID-19 pandemic with the aim of optimizing preventive and interventional measures \cite{Flaxman2020,Giordano2020,Gatto2020}. Our model, herewith described, could establish a new very precise dynamic estimate of the infection evolution by accurate tracking of the reproduction number, giving in advance the real morbidity and lethality measures. These are extremely helpful parameters for setting up preparedness and responsiveness plans for control of possible recurrent waves of SARS-CoV-2 as well as of future emerging and pandemic infections. Moreover, it is these authors' opinion that the present  model could be a critical indicator for establishing sustainability of any health system and preventing its collapse. Finally, for future studies the nonparametric approach here developed can be incorporated also in models which have a broader scope than recovering the reproduction number and the number of infected.\\

\newpage

\centerline{\bf \huge Methods}
\medskip

We describe our nonparametric approach for 
contact rate estimation in the context of the SIR model
used to generate the results described in the paper.
This is done without loss of generality.
As also clear in the sequel, all the ideas and mathematical results here obtained 
then apply to any compartmental model, hence defining
the entire nonparametric class
illustrated in Fig. \ref{Fig1}.

\subsection*{Time-varying SIR} 

Consider, without loss of generality, a population  normalized to $1$.
According to SIR models, the notation $S(t)$ indicates the susceptible people (who can be infected), $I(t)$ denotes the infected people (who have been infected and are able to spread the infection), while $R(t)$ represents the removed people (who were infected but then either healed or died). Healed people acquire immunity, so that $S(t)$ is a decreasing function. In addition, susceptible people can be infected through dynamics depending on the number of contacts between infected and susceptible ones, i.e.
$$
\dot{S}(t)=-a(t)S(t)I(t), \ a(t)>0.
$$
Above, $a(t)$ is the contact rate which is strongly related to the adopted restraints. 
The function $I(t)$ can both increase, because of susceptible  who become infected, and decrease, because of healing and/or death. The decreasing rate is proportional to $I(t)$, i.e.
$$
\dot{I}(t)=a(t)S(t)I(t)-bI(t), \ b>0.
$$
Finally, $R(t)$ increases accordingly to the healing/death rate, i.e.
$$
\dot{R}(t)=bI(t).
$$
In the above equations, $b$ describes average time for healing/death.  In absence of effective cures it can be assumed independent of time. 
We are only interested in positive solutions, i.e. $S(t),I(t),R(t) \ge 0$, which are necessarily bounded since
\begin{equation}
S(t)+I(t)+R(t)=1, \ \forall t \in {\mathbb R}.
\label{LIMITE}
\end{equation}

By defining 
$q(t):=\frac{b(t)}{a(t)}>0$, which corresponds to the inverse of the reproduction number
$\gamma(t)$, it easily follows that
$$
\begin{array}{lcl}
\frac{d}{dt} \{ I(t)+S(t)-q(t)\ln[S(t)]\}&=&-\dot{q}(t)\ln[S(t)] \ \Rightarrow \cr
I(t)+S(t)-q(t)\ln[S(t)]&=&1-\int_{-\infty}^t \ \dot{q}(\tau)\ln[S(\tau)] d\tau.
\end{array}
$$
The integral on the lhs is evaluated for $t \rightarrow -\infty$, when no infected and, hence, no removed are present.
So, one has $S(-\infty)=1$ and $I(-\infty)=R(-\infty)=0$. By defining
\begin{equation}
\delta(t):=-\int_{-\infty}^t \ \dot{q}(\tau) \ \ln S(\tau) d\tau
\label{DELTA}
\end{equation}
$I(t)$ and $R(t)$ become the following functions of $S(t)$: 
\begin{eqnarray}
\nonumber R(t)&=&g(S(t),\delta(t)):=-\delta(t)-q(t)\ln[S(t)], \\
\label{VARY} I(t)&=&f(S(t),\delta(t)):=1+\delta(t)-S(t)+q(t) \ln[S(t)].
\end{eqnarray}
This permits us to rewrite the differential equations in terms of $q(t)$ as
\begin{equation}
\dot{I}(t)=a(t)I(t)[S(t)-q(t)], \ \dot{S}(t)=-a(t)S(t)I(t).
\label {EQ DIFF}
\end{equation}

We have seen that the function $a(t)$ has to describe how the 
level of people social interactions evolves in time, accounting also for 
lockdowns effects.
Before the
lockdown's instant $t^*$, such a function is assumed constant.
Under these stationary assumptions, i.e. in absence of lockdowns, 
the function $q(t)$ is thus equal to the constant $q$ and one has
$\delta(t)=0$. This implies
\begin{eqnarray}
I(t)&=&f(S(t),0)=1-S(t)+q \ln[S(t)]\\ 
R(t)&=&g(S(t),0)=-q \ln[S(t)].
\label{INVARIANT}
\end{eqnarray}
Now, let the instant $t=0$ be the beginning of our experiment, 
with $S(0) \approx 1$.
The lockdown's instant is instead denoted by  
$t^*>0$ (corresponding to March 9, 2020, in Italy). 
Using (\ref{INVARIANT}), the following approximated relationship is obtained
\begin{equation}\label{ApproxSI}
S(0) \approx 1+\frac{I(0)}{q-1}.
\end{equation}

\subsection*{A parametric class of time-varying SIR}

We start introducing a parametric class of time-varying SIR models
instrumental for the building of the nonparametric approach. 
Our data model is
\begin{eqnarray*}
\dot{S}(t)&=&-a(t)S(t)I(t) \\
\dot{I}(t)&=&a(t)S(t)I(t)-bI(t) \\ 
\dot{R}(t)&=&bI(t)\\
y(t)&=&\frac{1}{H}I(t). 
\end{eqnarray*}
Note that the measurable output $y(t)$ is proportional to 
the number of infected people through the inverse of the 
unknown parameter $H$. 
The state dynamics then depend on the parameter $b$ and 
the time-varying contact rate $a(t)$.\\ 
Since the
system is assumed to be stationary before the lockdown's instant, 
initial conditions are 
\begin{eqnarray*}
S(0)&=&1+\frac{Hy(0)}{q(0)-1}, \quad q(0)=\frac{b}{a(0)}\\
I(0)&=&Hy(0)\\
R(0)&=&1-I(0)-S(0)
\end{eqnarray*}
where in the expression of $S(0)$ we have assumed exact
the approximation (\ref{ApproxSI}).\\

While $a(t)$ is constant 
before the
lockdown's instant $t^*$, next we assume that it has a discontinuity in $t^*$. 
Then, during the restrictions, its value could still decrease
e.g. due to people's growing awareness of infection risk
or because restrictions can be further strengthened after the first lockdown. 
One simple parametric
time-course for $a(t)$ is given by 
\begin{eqnarray}\label{hdn} 
\quad \qquad a(t)=  \left\{ \begin{array}{cl}  
    a_1  &  \quad  \mbox{if} \ \  t< t^* \\ 
    a_2 e^{-c(t-t^*)}   &   \quad \mbox{if} \ \  t^* \leq t \leq t_{end}  
\end{array} \right.
\end{eqnarray}
where  $t_{end}$ denotes the end of the lockdown (May 18, 2020, in Italy).
Note that  $a(t)$ would tend to zero if 
the lockdown would never end ($t_{end}=+\infty$).\\
The above model does not follow the paradigm depicted in Fig. \ref{Fig1}
because the contact rate is not described through an infinite-dimensional model. 
It depends on parameters which are the components of the
following finite-dimensional parameter vector
$$
\theta=\left[ a_1 \  a_2 \ b \ c \ H\right].
$$
For our future developments, we need  to prove that $\theta$ is globally identifiable from data, i.e.
it can be reconstructed under the ideal assumption of knowledge of the entire 
output trajectory $y(t)$. Using differential algebra tools, e.g. see \cite{Saccomani2003}, 
the system leads to the following characteristic set: 
$$
\frac{\dot{a}(t)}{H}\dot{y}(t)y(t) - \frac{a(t)}{H} \ddot{y}(t) y(t)  +\frac{a(t)}{H} \dot{y}^2(t) + a^2(t)\dot{y}(t)y^2(t) - \dot{y}(t)y(t) \frac{ba(t)}{H}. 
$$
If $t < t^*$, one has $a(t)=a_1$ and $\dot{a}(t)=0$ so that the coefficients of the characteristic set become 
$$
\frac{a_1}{H}, a_1^2, \frac{ba_1}{H}.
$$
Hence, since all the three parameters are known to be positive,  
the values of $a_1,H,b$ can be univocally determined.
If $t \geq t^*$, one has $a(t)=a_2e^{-c\tau},\dot{a}(t)=-ca_2e^{-c\tau}$. If $\tau_1=t_1-t^*$ 
and $\tau_2=t_2-t^*$  are two distinct and known time-instants, the characteristic set 
permits e.g. to reconstruct
$$
\frac{a_2e^{-c\tau_1}}{H},\frac{a_2e^{-c\tau_2}}{H}
$$ 
and this fact, combined with the knowledge of $H$, permits to achieve also $a_2$ and $c$.\\
Having shown that such parametric time-varying SIR is globally identifiable, 
we can use least squares to estimate $\theta$.
Let $y_{\theta}(t_i)$  denote the output of the SIR model as a function of the unknown parameter vector.
Then, two estimators are now considered. 
The first one uses a subclass of models defined 
by imposing $c=0$, making the contact rate description equal to
(\ref{PiecewiseConstant}). The resulting estimator is
$$
\hat{\theta} = \arg \min_{\theta \ s.t. \ c=0}  \sum_i (y(t_i) - y_{\theta}(t_i))^2
$$
and defines exactly the parametric estimates reported with dotted lines in Figs. \ref{Fig4} and \ref{Fig5}.\\
The other estimator exploits the entire class and is thus given by
$$
\hat{\theta} = \arg \min_{\theta}  \sum_i (y(t_i) - y_{\theta}(t_i))^2
$$
Using intensive care data in Lombardy, the estimates of the components of $\theta$
turn out 
$$
\hat{a}_1=0.27, \ \hat{a}_2=0.19, \ \hat{b=0.076}, \ \hat{c}=0.011, \  \hat{H}=1467.8.
$$
while, assuming Gaussian noise, the maximum likelihood estimate of the noise variance 
is $\hat{\sigma}^2$=3.5e-12. This corresponds to a standard deviation equal to 18.8 on the intensive
care data not normalized w.r.t. whole population in Lombardy which were shown in the right panel of Fig. \ref{Fig1abc}.
Thus, the estimate of $a(t)$ for $t\geq t^*$ is $0.19e^{-0.011}t$ and provides a 
first hint as how the contact rate decreased during the lockdown in Lombardy.
But it is questionable if a mono-exponential is suited to describe a so complex phenomenon.
For this reason, in the next section this simple model will be will be generalized  through 
nonparametric arguments.

\subsection*{Nonparametric model of the contact rate} 

We will assume that $a(t)$ belongs to a special class of Hilbert spaces $\mathcal{H}$ 
called Reproducing kernel Hilbert spaces (RKHSs)  \cite{Aronszajn50,Bergman50}.
To introduce them, recall that, if $\mathcal{X}$ denotes the function domain,  
$K:\mathcal{X}\times \mathcal{X} \rightarrow \mathbb{R}$ is called \emph{positive definite kernel} if, for any finite natural number $p$, it holds that
$$
\sum_{i=1}^{p}\sum_{j=1}^{p}c_ic_j K(x_i,x_j) \geq 0, \quad \forall (x_k,c_k) \in \left(\mathcal{X},\mathbb{R}\right), \quad k=1,\ldots, p.
$$
One can then prove that any RKHS is in one-to-one correspondence  
with a positive definite kernel and inherits the properties of the kernel,
e.g. continuous kernels induce spaces of continuous functions.
For our developments, the following fact is also important.  
Given a kernel $K$, the \emph{kernel section} $K_x$ centered at $x$ is the function
$\mathcal{X} \rightarrow \mathbb{R}$ defined by
$$
K_x(y) = K(x,y) \quad  \forall y \in \mathcal{X}.
$$
Then, one has that any function in $\mathcal{H}$ is a linear combination of a possibly infinite number of kernel sections
\cite{Cucker01}.\\
The question is now which RKHS can be conveniently introduced as hypothesis space for $a(t)$.
During a lockdown, this function is expected to have a smooth decay as time progresses.
We can then consider the so called first-order stable spline kernel defined by
\begin{equation}\label{InfDimSSker}
K(t,\tau) =\lambda e^{-\alpha \max{(t,\tau)}}, \quad 0 < \alpha <1, \ \lambda \geq 0
\end{equation}
which was originally introduced in the literature 
to describe impulse responses of stable systems \cite{Pillonetto:10a}.
It depends on the positive scale factor $\lambda$ and the
scalar $\alpha$ which regulates the 
decay rate of the functions contained in the associated RKHS.
We will fix these two parameters by exploiting
the estimates of the mono-exponential decay
 obtained in the previous section.
 In particular, we set $\lambda=\hat{a}_2^2$ and 
 $\alpha=2\hat{c}$. 
Thinking also of the Bayesian interpretation of regularization,
where the kernel is seen as a covariance \cite{Rasmussen},
this makes our space in some sense centred around 
exponentials of amplitude $\hat{a}_2$ and decay rate $\hat{c}$.
This fully defines the kernel and, hence, the associated RKHS $\mathcal{H}$.
It can be proved that such stable spline space is infinite-dimensional and able to
approximate any continuous map. 
Our nonparametric model for $a(t)$ is then defined by
\begin{eqnarray}\label{ContactIDmodel} 
\quad \qquad a(t)=  \left\{ \begin{array}{cl}  
    a  &  \quad  \mbox{if} \ \  t< t^* \\ 
    f(t-t^*) \ \mbox{with} \ f \in \mathcal{H} &   \quad \mbox{if} \ \  t^* \leq t \leq t_{end}  
\end{array} \right.
\end{eqnarray}
So, the overall model now follows the paradigm in Fig. \ref{Fig1} with $\theta=[a \ b \ H]$
and the $a(t)$  defined by $a$ and $f \in \mathcal{H}$.\\
Estimation of $f$ and $\theta$ is however ill-posed. This problem is circumvented using
regularization in $\mathcal{H}$ with penalty term defined by the RKHS norm $\| \cdot \|_{\mathcal{H}}$.
Specifically, letting $y_{f,\theta}(t_i)$  be the output of the SIR model as a function of $f$ and $\theta$, 
our estimator is given by 
\begin{equation}\label{IDproblem}
(\hat{f},\hat{\theta}) = \arg \min_{f \in \mathcal{H},\theta \in \mathbb{R}^3}  \sum_i \frac{(y(t_i) - y_{f,\theta}(t_i))^2}{\hat{\sigma}^2} + \| f \|_{\mathcal{H}}^2
\end{equation}
where $\hat{\sigma}^2$ is the  maximum likelihood estimate of the noise variance already 
mentioned in the previous section. The objective in (\ref{IDproblem}) includes two different components.
The first one is a quadratic loss and penalizes values of $\theta$ and $f$ associated to compartmental models
unable to well describe the observational data. The second one is the regularizer, defined
by the RKHS norm, which restores well-posedness. It excludes non plausible solutions for the contact rate $a(t)$,
e.g. defined by too irregular temporal profiles of $f(t)$.   
The problem thus corresponds to a nonlinear version of a regularization network \cite{Poggio,Scholkopf01b}.
Its solution exists since, according to the results in \cite{PilAutoNum08},
optimization can be restricted to a compact set of the continuous functions
equipped with the sup-norm where 
the map  $y_{f,\theta}$ is continuous (see also Appendix of \cite{PillSacc}) 
and the regularizer is lower semicontinuous.\\
However, differently from the classical machine learning problems 
where $f$ is linearly related to data, the nonlinearities 
present in our compartmental model makes the solution (\ref{IDproblem}) 
not available in closed-form. To compute it, the following 
strategy has been then adopted. By fixing an integer $M$, 
we define the following representation
$f(t)= \sum_i^{M} c_i K_{t_i}(t)$ given by sum of kernel sections over a uniform grid of temporal instants $t_i$.
Next, a sequence of solutions  of the problem (\ref{IDproblem}), 
with the objective restricted to these finite-dimensional subspaces of dimension $M$, 
is obtained for increasing values of $M$. This is done until 
reaching convergence (which is guaranteed still exploiting the results in \cite{PilAutoNum08}).\\
The stable spline kernel  is useful to describe the contact rate during the lockdown,
since it embeds information on smooth decay of the reproduction number. 
Since $\gamma(t)$ after the end of the lockdown is not expected to decay, and could also increase,
to obtain the results depicted in Fig. \ref{Fig9} we have used the Laplacian kernel \cite{Scholkopf01b}
\begin{equation}\label{InfDimLapker}
K(t,\tau) =\lambda e^{-\frac{| t-\tau |}{\eta}}, \quad \lambda,\eta > 0
\end{equation}
which embeds information only on continuity of the time-course.
System initial conditions at the end of the lockdown 
are set to the estimates obtained by the procedure reported above.
Next, a parametric model with constant contact rate $a(t)$  is fitted to data,
obtaining also a recalibration
of the noise variance, and 
the scale factor $\lambda$ is set to its squared value. The kernel width $\eta$
is then estimated through the concept of Bayesian evidence, exploiting the stochastic interpretation of 
(\ref{IDproblem}), as also discussed in the next paragraph, 
and using the Laplace approximation to compute the model posterior probability \cite{Raftery}.\\
Finally, to complement the estimates with confidence intervals, a Bayesian framework has been adopted
resorting to the stochastic interpretation of regularization and the duality between RKHSs and Gaussian processes
\cite{Rasmussen}. 
 The noise affecting the data is assumed to be Gaussian. The components of $\theta$, and also the noise variance, 
 are seen as mutually independent 
 random variables and are assigned poorly informative prior distributions, in practice including only nonnegativity information. 
The contact rate for $t \geq t^*$ is then seen as a Gaussian process
defined by $f(t)= \sum_i^{M} c_i K_{t_i}(t)$. Here, the $c_i$ are the components of the
zero-mean Gaussian vector $c$ whose covariance matrix is the inverse
of $\bar{K} \in \mathbb{R}^{M \times M}$ with $(i,j)$ entry given by $K(t_i,t_j)$.
In this way, the function $f(t)$ sampled on the $t_i$ is indeed Gaussian with covariance matrix $\bar{K}$.
Markov chain Monte Carlo has been then used to reconstruct the posterior in sampled form \cite{Gilks}.
In particular, a random walk Metropolis  has been implemented. Covariances of the increments
have been tuned through a pilot analysis 
to obtain an acceptance rate around  $30\%$ and then 4 million iterations have been performed.

\subsection*{Data availability}  \
All data used in this manuscript are publicly
available 
at https://github.com/pcm-dpc/COVID-19. 

\bibliography{BiblioRegBook} 

\end{document}